\documentclass[12pt,preprint]{aastex}

\usepackage{natbib}

\newcommand{\pref}{\protect\ref}
\newcommand{\solrad}{\ifmmode{R}_{\rm S}\else${R}_{\rm S}$\fi}
\newcommand{\solmas}{\ifmmode{M}_{\rm S}\else${M}_{\rm S}$\fi}

\newcommand{\ctn}{\ifmmode\kappa\else$\kappa$\fi}

\newcommand{\flxu}{$\,$ergs$\,$cm$^{-2}\,$s$^{-1}$}
\newcommand{\velu}{$\,$km$\,$s$^{-1}$}

\newcommand{\term}[2]{\mbox{$\,^{#1}{\rm #2}$}}
\def\term#1 #2/{\mbox{$\,^{#1}{\rm #2}$}}

\newcommand\lta { \mathrel {\hbox to 0pt {\lower 3.7pt \hbox{$\sim$}
      \hss} \raise 1.7pt \hbox{$<$}}}
\newcommand\gta { \mathrel {\hbox to 0pt {\lower 3.7pt \hbox{$\sim$}
      \hss} \raise 1.7pt \hbox{$>$}}}

\newcommand{\philemail}{judge@ucar.edu}
\newcommand{\luciaemail}{lucia.kleint@fhnw.ch}
\newcommand{\alinaemail}{alina.donea@monash.edu}
\newcommand{\albertoemail}{asdalda@stanford.edu}
\newcommand{\lyndsayemail}{lyndsay.fletcher@glasgow.ac.uk}

\newcommand\tabone{
\protect\begin{deluxetable}{llll}
\protect\label{tab:tabone}
\renewcommand{\baselinestretch}{1.2}
\tablecaption{Properties of the acoustic source and X1 flare of 29 March 2014}
\tablehead{Property & Instrument & timing UT & position }
\startdata
Impulsive phase start & RHESSI 30-70 keV & 17:45  \\
Impulsive phase peak & RHESSI 30-70 keV & 17:47:16 &  519.7, 263.2\\
Acoustic source peak & HMI 5.5 mHz & 17:48 & 518.5, 264.0 \\
                     & HMI 6 mHz & 17:51 \\
IR continuum peak    & FIRS & $\approx$ 17:46:10 & $\approx$ 520, 263\\
IR \ion{Si}{1} core peak & FIRS & $\approx$ 17:46:37 & $\approx$ 519.7, 263.5\\\\
IR \ion{He}{1} core peak & FIRS & $\approx$ 17:46:10 & $\approx$ 521, 263\\
\enddata 
\tablecomments{ 
Note that FIRS is not an imaging instrument, the slit was scanned across
the solar surface (see Figure~\pref{fig:context}).  }
\end{deluxetable}
}

\newcommand{\figcontext}{
\begin{figure}[] 
\epsscale{1.0}
\plotone{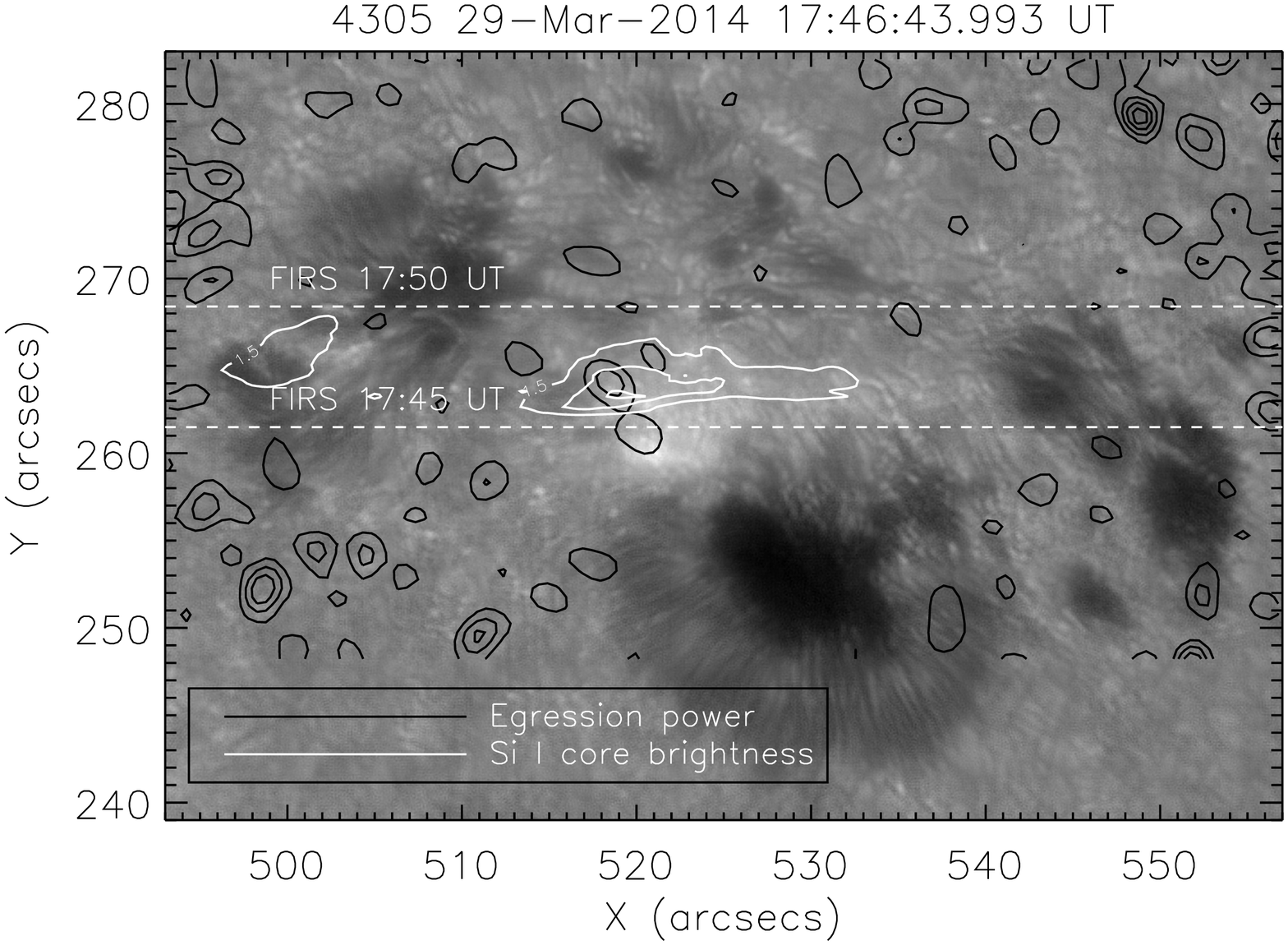}  
\caption{\label{fig:context} A G-band image centered near 430.5 nm
  from the Dunn Solar Telescope is shown along with two sets of
  contours.  The black contours are of egression power, $|H_{+}(r, t)|^{2}$, 
  referred to in the text.  The white
  contours are of core intensity of the 1082.7 nm line of \ion{Si}{1}.
  The FIRS data were acquired between 17:40:12 and 18:01:39 UT by
  stepping the slit (horizontal in the figure) in the Y-direction (S-N).
  The positions of the slit during the flare are shown as dashed
  lines.  The absolute heliographic coordinates of our ground-based data 
are accurate to 1 second of arc, as obtained by co-alignment with a continuum image from HMI obtained at 17:50:00 UT. }
\end{figure}
}

\newcommand{\figevolone}{
\begin{figure}[] 
\epsscale{1.0}
\plotone{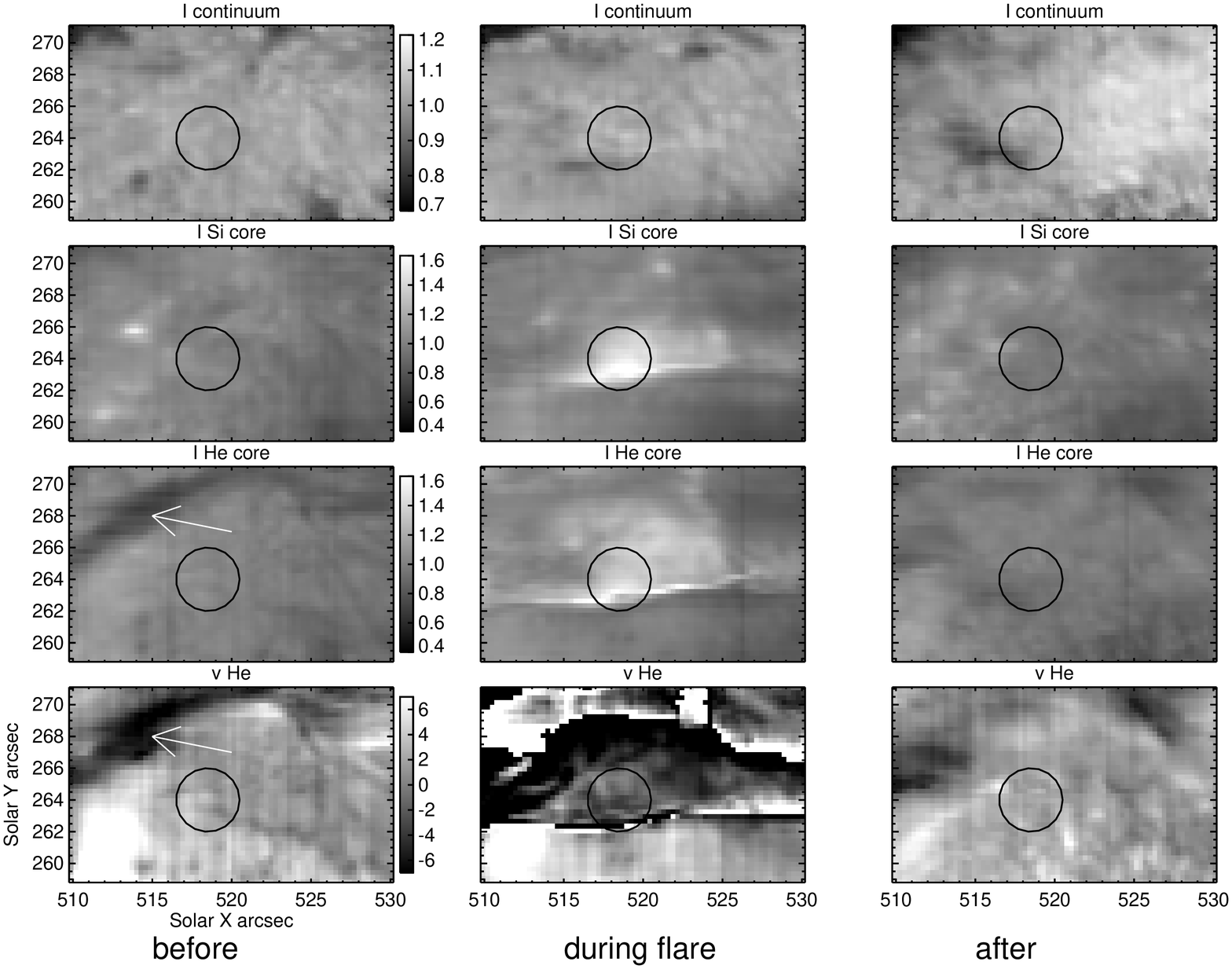}  
\caption{\label{fig:evolone} ``Quick-look'' raster images of thermal data
  obtained using the FIRS instrument.  The columns, from left to
  right, show data from 16:29:26, 17:40:06 (during the impulsive phase
  of the flare) and 18:30:13 respectively.  The top panels show
  continuum intensity averaged over 30 pixels on the blue side of the
  Si I 1028.7 nm line.  The next two rows shows the intensity at the
  cores of the Si I and He I 1083.0 nm lines.  Row 4 shows Doppler
  shifts of the \ion{He}{1} multiplet in units of \velu{}.  The flare began around 17:45
  UT, at about the time the FIRS slit crossed the flare ribbon seen in
  the Si I and He I core intensities (middle column, 
refer also to Figure~\pref{fig:context}).
  The circle shows the location of the ``acoustic source'' associated
  with the flare in the field of view shown. The arrows mark the erupting 
filament seen in \ion{He}{1}. }
\end{figure}
}

\newcommand{\figevoltwo}{
\begin{figure}[] 
\epsscale{1.0}
\plotone{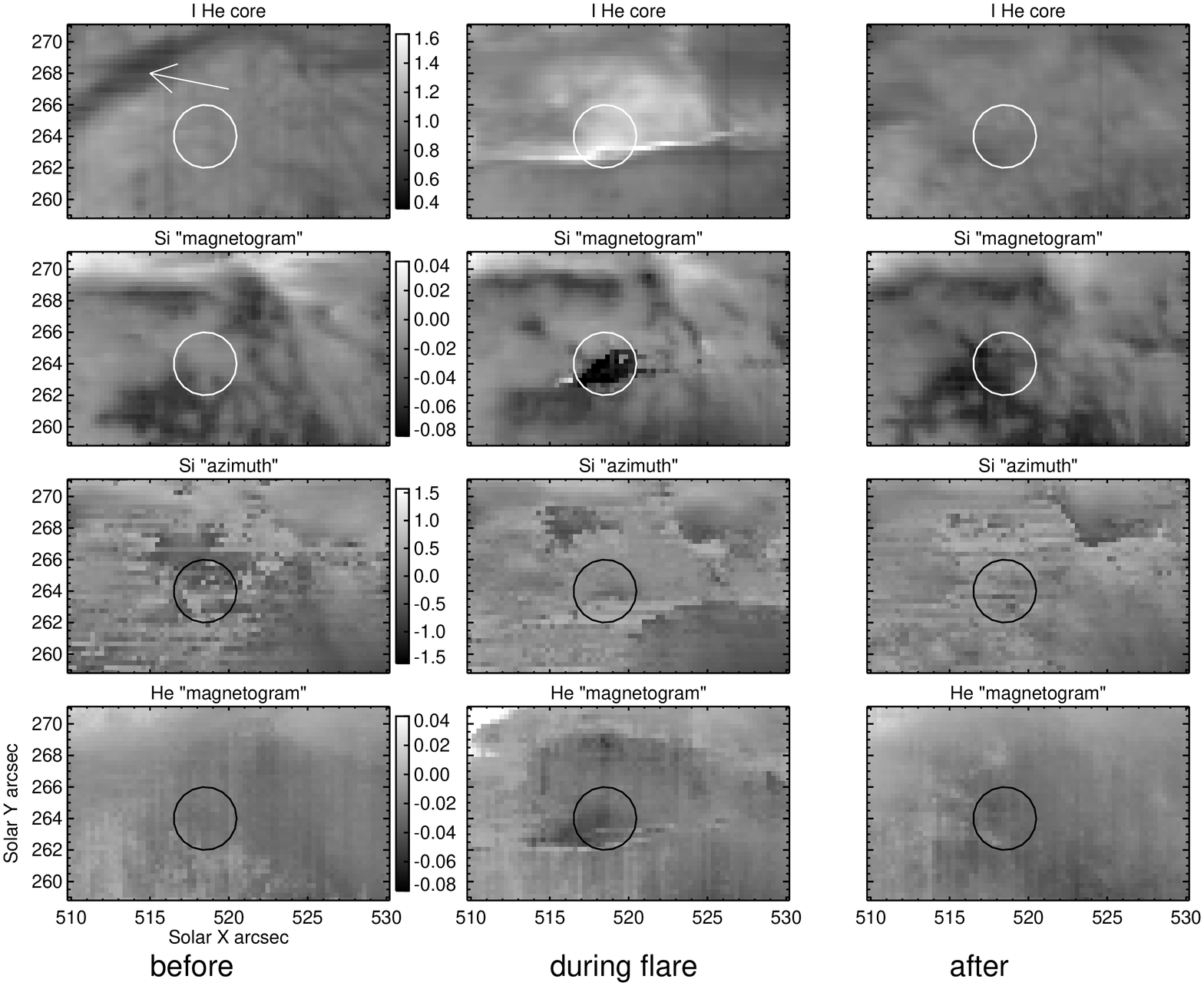}  
\caption{\label{fig:evoltwo} Quick-look raster images relevant to
  magnetic fields obtained using the FIRS instrument.  The top panels
  show the intensity near the core of the He I 1083.0 nm lines
from Figure~\pref{fig:evolone}.  Rows 2
  and 3 show ``magnetograms'' and ``azimuths'' of the magnetic field
  (see text), of the \ion{Si}{1} line.  Rows 4 and 5 show similar data but
  for the \ion{He}{1} multiplet.  Care should be taken not to interpret the
  crude magnetic data in terms of magnetic field especially during the
  flare, which occurred during the scan shown in the second column.  }
\end{figure}
}

\newcommand{\fignlte}{
\begin{figure}[] 
\epsscale{1.0}
\plotone{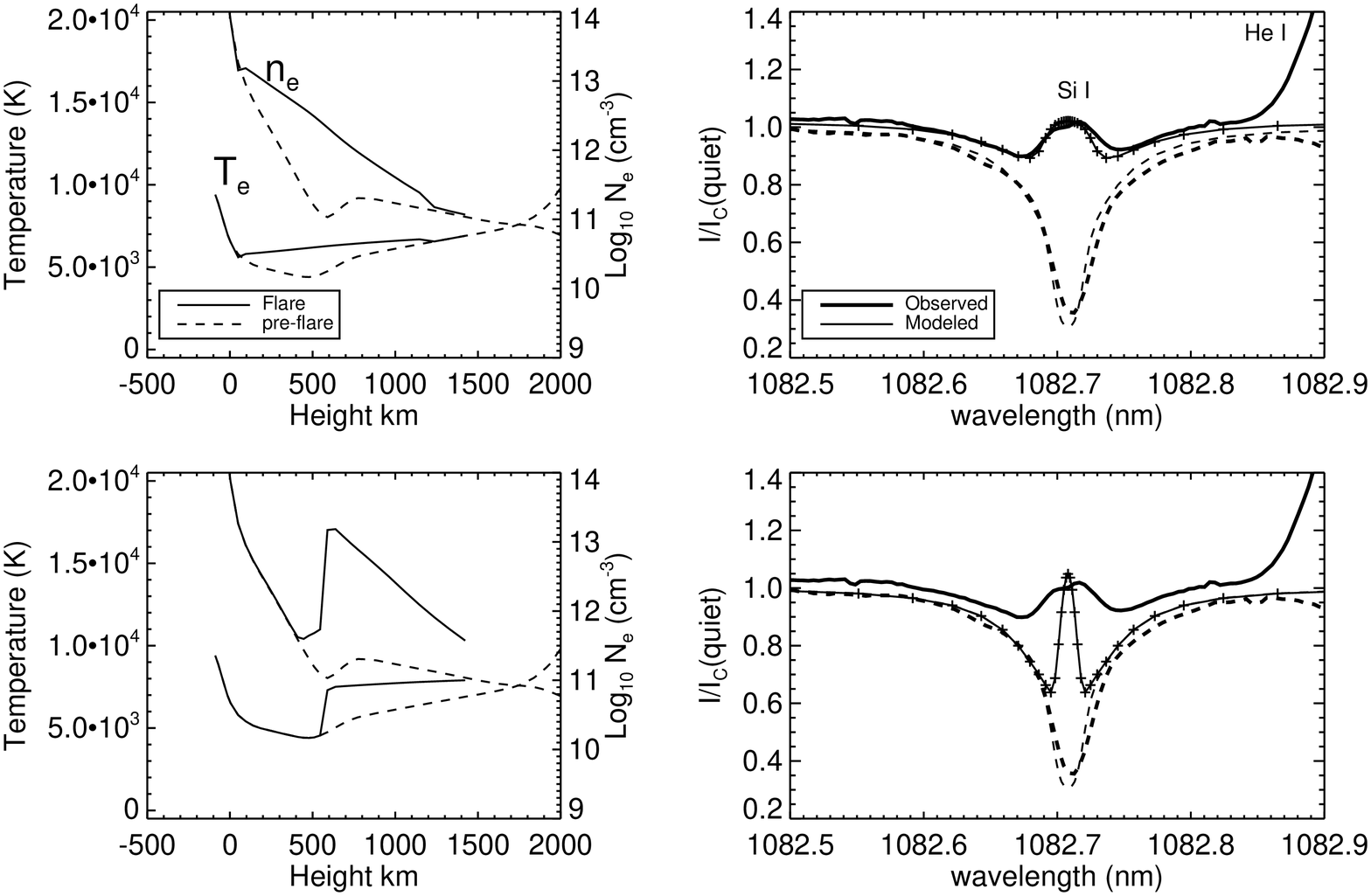}  
\caption{\label{fig:nlte}
The results of nLTE calculations are shown.  The atmospheric
structure is shown in the left panels, emergent intensities are shown in the 
right panels, for a heliocentric angle with cosine 
of 0.95.   Two ``flare models'' are shown, exploring the  depths and magnitudes of
enhanced temperatures in the flaring atmosphere.    Typical observed profiles of the flare ribbon
and a non-flaring region from the same exposure are shown as thin solid and dotted lines 
respectively.   The upper panels show a ``deep penetration'' model.  The lower 
panels allow penetration to a depth near 600 km such that the \ion{Si}{1} line is in emission,
but is a shallow penetration model.  Notice that the shallow model 
predicts no detectable increase in 
continuum
emission and it produces a line emission core that is in qualitative disagreement with the
data.
}
\end{figure}
}

\newcommand{\figstokes}{
\begin{figure}[] 
\epsscale{1.0}
\plotone{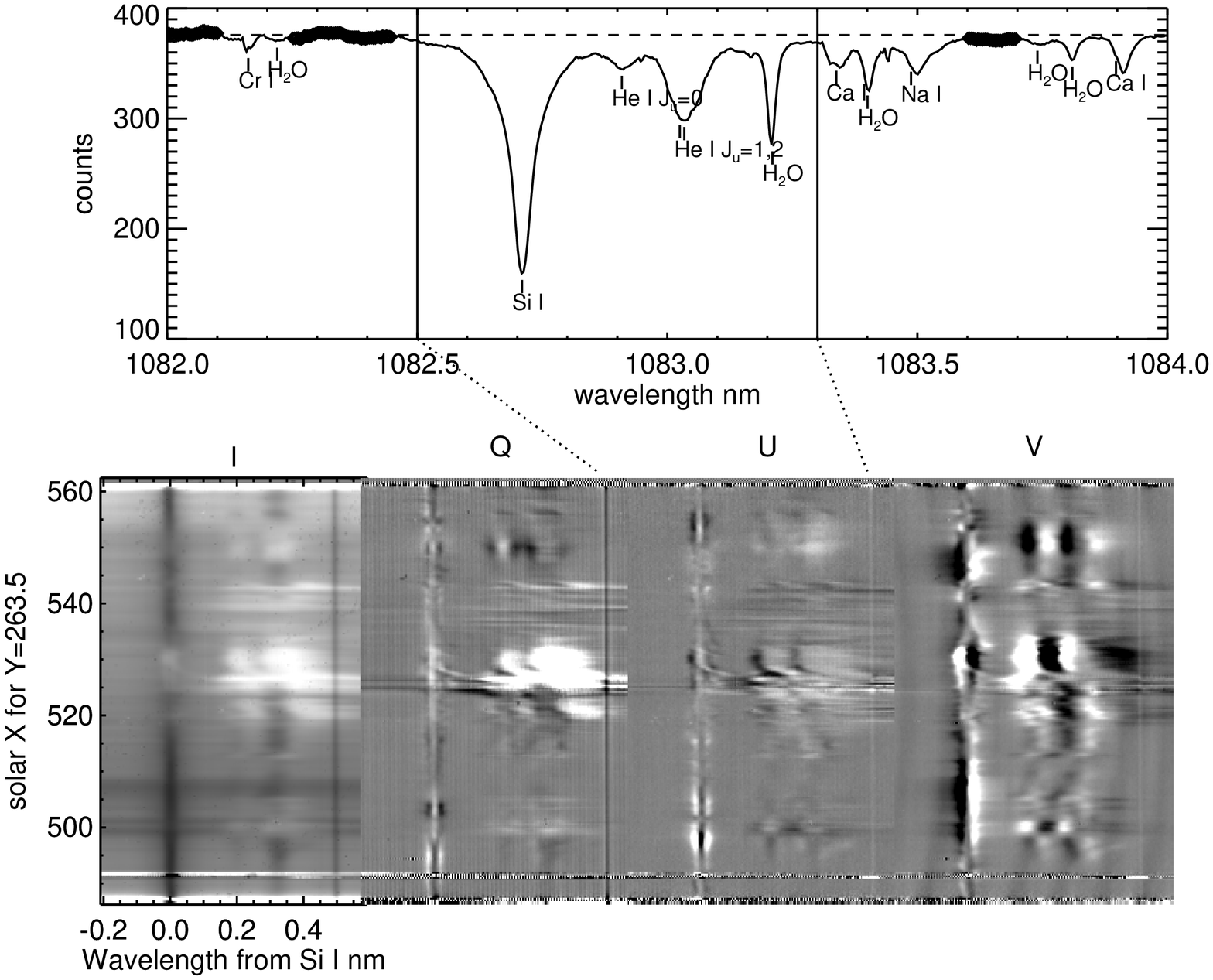}  
\caption{\label{fig:stokes}
The upper panel shows the median intensity spectrum from the spectral scan 
obtained by FIRS started at 17:40:06 UT.   Line identifications are marked, the 
${\rm H_2O}$ features are all telluric and should be polarization-free.  Those
wavelengths used to derive the continuum intensity are marked on the spectrum 
with symbols.  
The four lower panels show Stokes $I,Q,U$ and $V$ spectra taken from the 30th
scan obtained through the flare footpoint beginning at 17:46:29 UT, as a function of
wavelength and position on the Sun.   The dotted lines show the wavelengths
plotted on the $U$ image.  The images for the 
$QUV$ images are clipped at $\pm 2\times10^{-2} I_C$.  
The \ion{Si}{1} line is in emission near 
$X\approx530$.  The $QUV$ profiles of this line are as expected from 
Zeeman-induced polarization, even when the line is in emission. The \ion{He}{1} 
$QU$ and $V$ profiles are peculiar and to some degree reflect the systematic errors 
of ``crosstalk'' ($QU$ have some characteristics of $I$ and $V$).  Some spectral fringes are 
visible, particularly in the $Q$ and $V$ images.
}
\end{figure}
}

\newcommand{\figinvone}{
\begin{figure}[] 
\epsscale{1.0}
\plotone{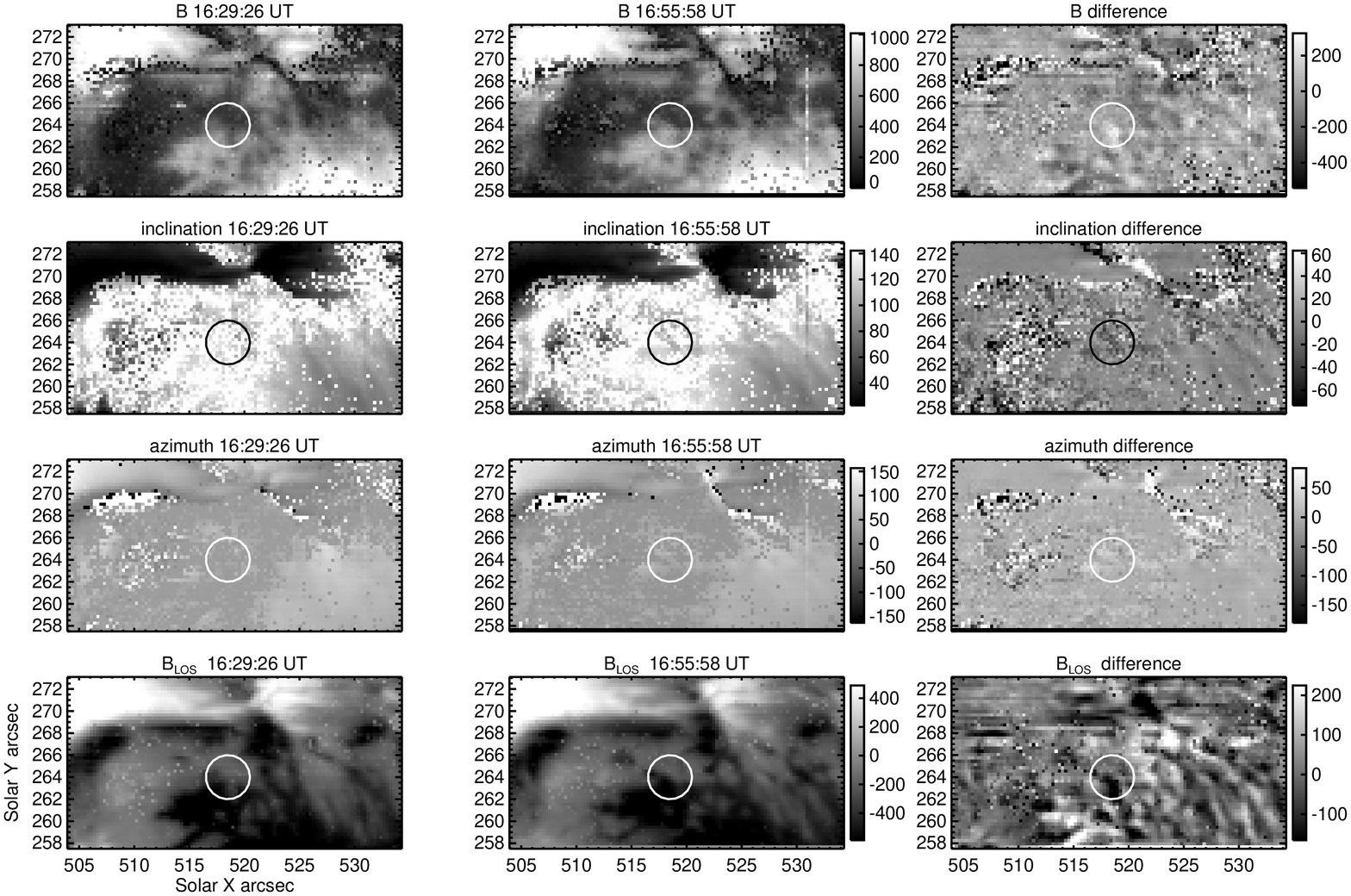}  
\caption{\label{fig:invone} Magnetic field properties determined from the scans 
obtained before the flare, begun at 16:29:26 and 16:55:58 UT, using the code MELANIE.   The azimuth is
measured in the E-W direction, the sign of the azimuth is not determined.  The angles 
are measured in the local solar frame (E-W and local vertical reference directions).  The circle shows 
the center of the acoustic source.   The rightmost column shows differenced data.  
}
\end{figure}
}

\newcommand{\figinvtwo}{
\begin{figure}[] 
\epsscale{1.0}
\plotone{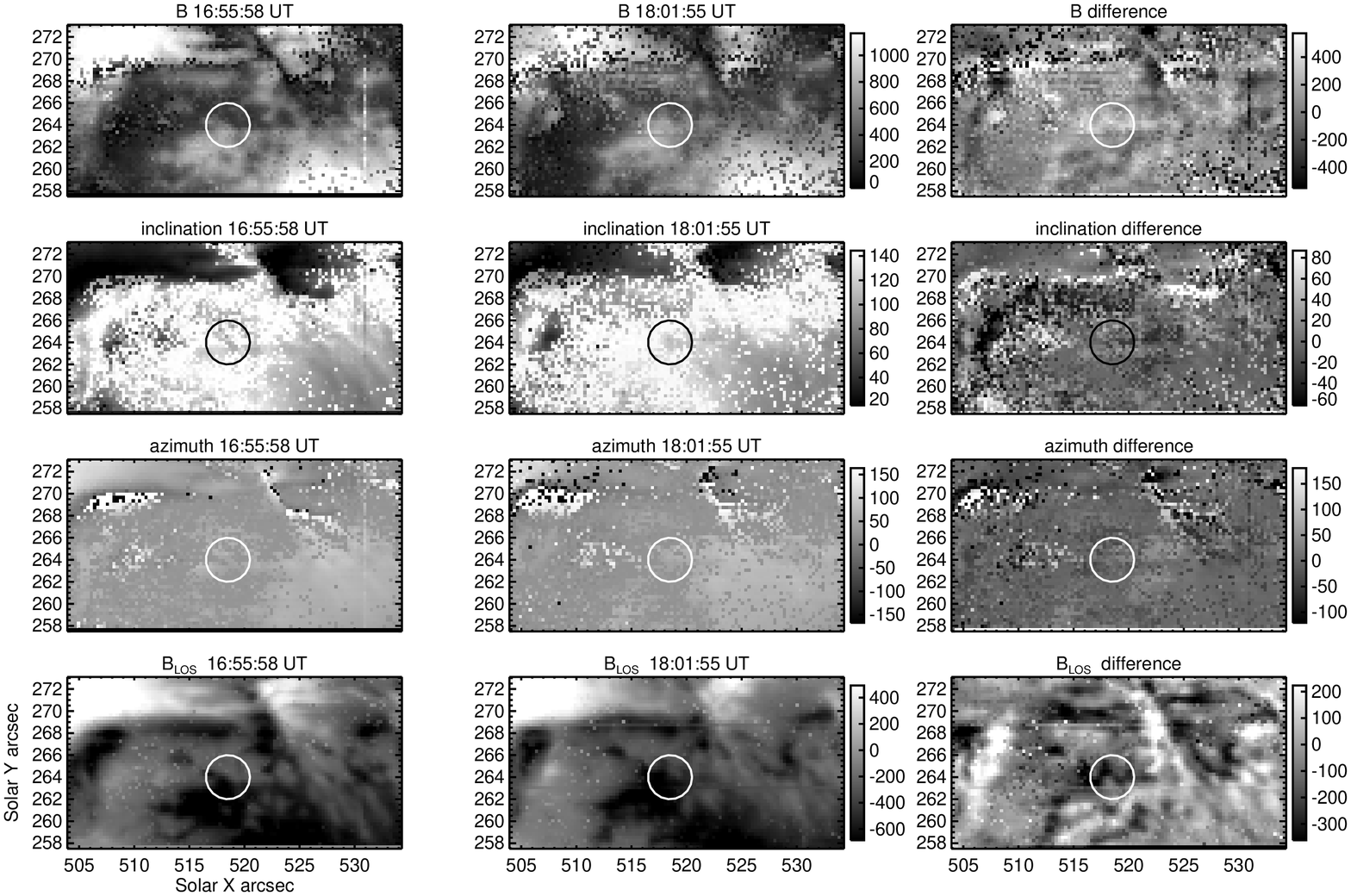}  
\caption{\label{fig:invtwo} Magnetic field properties determined from the scans 
obtained before (16:55:58 UT) 
and after (18:01:55 UT) the flare, shown as in Figure~\pref{fig:invone}. }
\end{figure}
}

\newcommand{\figcont}{
\begin{figure}[] 
\epsscale{1.0}
\plotone{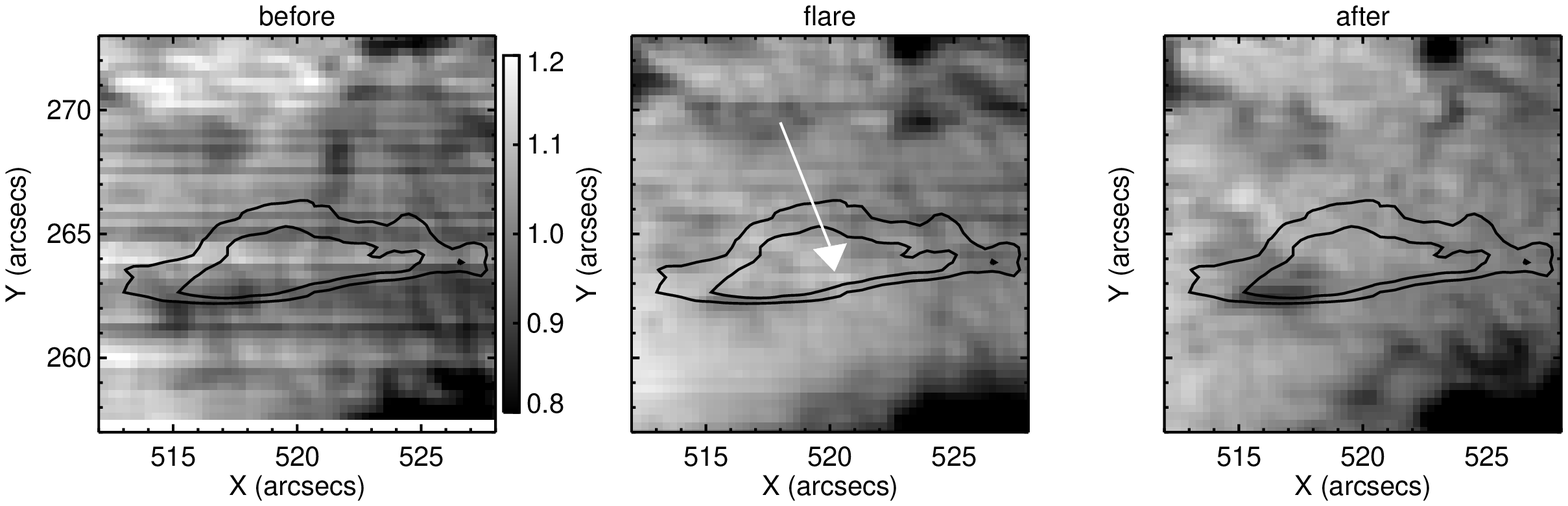}  
\caption{\label{fig:continuum} Continuum intensity data are shown
from the three scans immediately before, during and after the X1 flare
impulsive phase.   The contours show the intensity from the core of the \ion{Si}{1} line.  
Enhanced intensity in the continuum is seen as the central part of one row
in the image 
the middle panel, marked by an arrow.  Its amplitude is $4\pm1\%$ of the neighboring intensities. 
}
\end{figure}
}

\shortauthors{Judge et al.}
\shorttitle{acoustic flare sources}

\slugcomment{}

\begin{document}

%
%

\title{On the Origin of a Sunquake during the 29 March 2014 X1 Flare}
\author{  
Philip G. Judge}
\affil{High Altitude Observatory,
       National Center for Atmospheric Research\footnote{The National %
       Center for Atmospheric Research is sponsored by the %
       National Science Foundation},
       P.O.~Box 3000, Boulder CO~80307-3000, USA; \philemail}


\author{Lucia Kleint}
\affil{ Institute of 4D Technologies, University of
  Applied Sciences and Arts Northwestern Switzerland, 5210 Windisch,
  Switzerland; \luciaemail}

\author{Alina Donea} \affil{Center for Astrophysics, School of
  Mathematical Science, Monash University, Victoria 3800, Australia; \alinaemail}

\author{ Alberto Sainz Dalda}
\affil{Stanford-Lockheed Institute for Space Research, 
Stanford University, HEPL, 466 Via Ortega, Stanford, CA 94305, USA; \albertoemail}

\and
\author{Lyndsay Fletcher}
\affil{
SUPA, School of Physics and Astronomy, University of Glasgow, Glasgow G12 8QQ, UK; \lyndsayemail}

%
%

\begin{abstract}

  Helioseismic data from the HMI instrument have revealed a sunquake
  associated with the X1 flare SOL2014-03-29T17:48 in active region
  NOAA 12017.  We try to discover if acoustic-like impulses or actions
  of the Lorentz force caused the sunquake.  We analyze
  spectro\-polarimetric data obtained with the Facility Infrared
  Spectrometer (FIRS) at the Dunn Solar Telescope (DST).  Fortuitously
  the FIRS slit crossed the flare kernel close to the acoustic source,
  during the impulsive phase.  The infrared FIRS data remain
  unsaturated throughout the flare.  Stokes profiles of lines of Si I
  1082.7\,nm and He I 1083.0\,nm are analyzed.  At the flare
  footpoint, the \ion{Si}{1} 1082.7 nm core intensity increases by a factor
  of several, the IR continuum increases by $4\pm1\%$.  Remarkably, the \ion{Si}{1}
  core resembles the classical \ion{Ca}{2} $K$ line's self-reversed
  profile.  With nLTE radiative models of H, C, Si and Fe these
  properties set the penetration depth of flare heating to 100 $\pm
  100$ km, i.e. photospheric layers.  Estimates of the non-magnetic
  energy flux are {\em at least} a factor of two less than the
  sunquake energy flux.  Milne-Eddington inversions of the \ion{Si}{1}
  line show that the local magnetic energy changes are also too small
  to drive the acoustic pulse.  Our work raises several questions:
  Have we ``missed'' the signature of downward energy propagation? Is
  it intermittent in time and/or non-local?  Does the 1-2 s
  photospheric radiative damping time discount compressive modes?
\end{abstract}


\section{Introduction}

Flares are among the most energetic phenomena in the solar system,
with well-known impacts on the Earth.  Beginning in the 1960s, it
became clear that the only option for storing the large amount of
energy for sudden release is the free energy associated with the
magnetic field threading the Sun's atmosphere.  According to the
standard model
\citep{Carmichael1964,Sturrock1966,Hirayama1974,Kopp+Pneuman1976},
flares start by magnetic reconnection in the tenuous coronal
plasma. Only here is the Alfv\'en speed sufficiently high to permit
rapid evolution.  Subsequently, downward directed energy in the forms
of accelerated particles, magneto-plasma waves, radiation and thermal
conduction deposit energy from above leading to bright ribbon-like
structures in the chromosphere and, during strong flares, in the
photosphere. Such temporarily heated structures (durations of minutes)
then evaporate plasma into the corona, leading to post-flare loops
that are bright in soft X rays and UV radiation on time scales of
hours.

 Local helioseismology has revealed flares which are accompanied by
 acoustic pulses (``sunquakes'') propagating below the visible surface.  The
 mechanisms by which the flare disturbance, originating high in the
 solar atmosphere, couples to interior modes is not known, there being
 several challenges.  Firstly, flares are difficult to observe,
 generally speaking, at both the necessarily small time and length
 scales associated with the initial energy release (impulsive phase).
 Secondly, most acoustic sources preferentially occur in the
 magnetically complex penumbrae of sunspots \citep[e.g.][section
   3.6]{Fletcher+others2011}.  Thirdly, the energy has to propagate
 through the 9 pressure scale heights of the
 poorly-constrained 
chromosphere. The chromosphere
of active regions appears to be as complex as sunspot penumbrae
   \citep{Judge2010m}. Measurements of magnetic fields there are
   difficult and rare (Navarro 2005a,b; Uitenbroek 2011).
   \nocite{Socas-Navarro2005a,Socas-Navarro2005b, Uitenbroek2011}

Progress on sunquakes has been significant.  For example,
\citet{Donea+Lindsey2005} have demonstrated that only a small
fraction, $\lta 10^{-3}$ of the flare energy, is needed to trigger a
seismic transient in the photosphere. How this happens it is not yet
understood \citep[see, e.g., discussions by
][]{Donea2011,Kosovichev2014}.  Recent space missions have vastly
improved our ability to understand the evolving photospheric magnetic
field, and significant steps have been taken towards understanding
changing magnetic fields and flares.  The lower solar atmosphere can
show stepwise changes in line-of-sight (LOS) magnetic field strength
\citep{Kosovichev+Zharkova1999} and shear \citep{Wang1992} during
flares.  \citet{Sudol+Harvey2005} observed a  LOS field
change in 15 X-class flares with a median of 90 G. 
Recent observers have found photospheric field and inclination changes
even during small B1 flares \citep{Murray+others2012}, using
vector spectro-polarimetric data from the SP instrument 
on the Hinode spacecraft.  

In this paper we relate acoustic sources found by
\citet{Donea+others2014} from data from the Helioseismic Magnetic
Imager (HMI) on the Solar Dynamics Observatory (SDO) spacecraft, to
measurements from the Facility Infrared Spectrometer
\cite[``FIRS''][]{Jaeggli2011} at the Dunn Solar Telescope of the National 
Solar Observatory in Sunspot,
New Mexico.  Table~1 lists some properties of the flare 
and acoustic source from \citet{Donea+others2014}.  While
many acoustic sources are present on the Sun with this intensity,
the spatial and temporal characteristics of this particular source 
mark it as generated by the flare.   
Changes in the
thermal and magnetic structure in the atmosphere are reflected in 
our Stokes polarimeter data through a photospheric \ion{Si}{1} line at
1082.7 nm and in the \ion{He}{1} 1083.0 nm multiplet formed near the
top of the chromosphere.

\section{Observations}
\label{sec:obs}

We made spectropolarimetric observations on March 29 2014 with FIRS
and the Imaging BIdimensional Spectropolarimeter (IBIS) instruments
at the DST.  The latter will be reported elsewhere.  In addition, we
acquired, every 60 seconds, bursts of data in G-band (430.5 nm) and
Ca~II (393.3 nm) narrow band filters for speckle reconstruction, and a
white light camera acquired rapid cadence images.  FIRS was used in a
single slit, dual-beam mode with a 40 micron wide slit, subtending 
an angular width of $0\farcs30$,  oriented close
to the E-W line on the Sun.  

The polarization modulation scheme was a four-state balanced scheme
with 125 ms exposures and a full cycle of 1.2 s, with 10 such cycles co-added by the
instrument at each scan position on the Sun.  This relatively
slow modulation, set by the need for the liquid crystal variable retarders to relax in response 
to voltage changes,  runs the risk of encoding light variations
entering the polarimeter due to residual seeing motion and/or solar
evolution into systematic errors called ``crosstalk''
\citep{Lites1987,Judge+others2004,Casini+deWijn+Judge2012}.
Crosstalk appears to be important in the \ion{He}{1} line during 
the flare as the tenuous chromospheric  plasma radiating the helium emission 
evolves on timescales comparable or faster than the 1.2s modulation cycle. 
No evidence is seen for such crosstalk in the photospheric \ion{Si}{1} line.

Throughout all our observations the count
rate remained in the linear regime of the IR detector (below 8000
ADU).  The solar image was scanned from S to N across the FIRS slit in
100 or 120 steps of $0\farcs3$ to produce images in four
spectropolarimetric states $S_i$ (linear combinations of $I,Q,U$ and $V$),
covering a spectral range from 1081.93 to 1085.01\,nm, and a spatial
area of $30$\arcsec or $36\arcsec\times75\arcsec$ for all scans reported
here.  The images have bin sizes of $0.3\arcsec$.  Five scans of the
slit across NOAA 12017 were begun at 16:29:26, 16:55:58, 17:40:06,
18:01:55, 18:30:13 UT.  The seeing was good enough for the adaptive optics system (AO)
to maintain a lock on the sunspot during the observing run.  
The peak flare emission is seen in the FIRS
data during the third scan begun at 17:40:06 UT.

Figure~\pref{fig:context} shows a G-band image with
contours superposed, showing (black) the egression power from 
the acoustic holography reported by \citet{Donea+others2014}, and (white)
the core intensity of the \ion{Si}{1} line at 1082.7 nm.  The G-band image
was aligned with a continuum image from the HMI instrument on the SDO
spacecraft obtained at 17:45:00 UT by eye, co-alignment uncertainties
are at most one arcsecond.  (The co-alignment accuracy is limited
by the fact that the FIRS scan was obtained under varying seeing
conditions and over a 20 minute scanning period).  Black contours show
the dominant local sources of power for waves traveling down into the
solar interior.  The white contours show the influence of heating
processes from the flare on the regions of formation of the \ion{Si}{1} line
in the Sun's atmosphere.  The G-band image is a composite, speckle
reconstructed image obtained at 17:46:44 UT.  The image shows a
diffuse brightening at these wavelengths centered near X=522, Y=260,
which is real flare emission, with an amplitude of $\approx 1.2-1.4$ times the 
non-flaring intensity, perhaps a component of the still poorly 
understood white light
emission.  None of these data are strictly contemporaneous, the FIRS data shown
were built of a raster scan that began at Y=254.5 at time 17:40:12 UT,
and ended at Y=284.2 at 18:01:39 UT.  The horizontal dashed lines show
the positions of the FIRS slit at 17:45 and 17:50 UT.

\subsection{Data Reduction}
\label{subsec:reduce}

The FIRS data were reduced using software originally developed by
\citet{Jaeggli2011} and modified by Tom Schad (private communication 2013).  
The reductions
followed standard procedures: correction for detector non-linearities;
subtraction of dark frames; division by flat fields; co-registration
of the two beams (including corrections for image rotation);
polarization calibration; de-modulation (conversion of linear
combinations of Stokes parameters to individual Stokes parameters).
Since the required polarization sensitivity is very high in
chromospheric lines \citep[e.g.][]{Uitenbroek2011}, special care is
needed in handling calibrations. Usual dark frames and flat fields
were acquired, and a gain linearization correction was applied to the data
using a curve from \cite{Jaeggli2011}.  We used flats obtained with a 
calibration lamp which is vignetted across the detector frame, in preference to
solar flats in which photospheric spectral lines are always present. 
This is because we analyze below the detailed profiles of the 
\ion{Si}{1} line at 1082.7 nm.

Residual fringes and some detector artifacts are present in
these data.   Fringes are a source of systematic
error.  Using careful corrections for flat fields we have reduced
fringing to $\lta 2\times10^{-3}I_c$ (peak-to-trough) where $I_C$ is the continuum
intensity, which is defined using wavelengths for each individual scan shown in 
Figure~\pref{fig:stokes}.    The wavelength scales of the spectra 
are determined using solar flat-field scans and solar photospheric absorption
lines.   The spectrograph was  stable at the level of 0.2 pixels in wavelength 
(0.004 nm) during the observations, equivalent to a Doppler shift of 0.2 \velu{}. 

It is important to note that
at infrared wavelengths, the enhancement of intensity during flares is
moderate, quite unlike the well-known enormous UV and X-ray enhancements.  All of the FIRS
data were obtained in the linear regime.

\subsection{Stokes line profiles}

Figure~\pref{fig:stokes} shows the mean intensity spectrum with
annotated spectral features and, in the lower panels, Stokes profiles
for $I,Q,U$ and $V$ from left to right.  The particular data shown in
the lower panels are from the 29th scan obtained through the
flare footpoint beginning at 17:46:29 UT, just as the flare was in the
impulsive phase as found from RHESSI data analysis
\citep{Donea+others2014}.  The line profiles of the photospheric \ion{Si}{1}
line are essentially consistent with polarization induced by the
Zeeman effect, with the possible exception of those seen in the flare
kernel.  On the other hand, while the \ion{He}{1} linear polarization ($Q,U$) profiles
might initially appear to be of solar origin, the result of atomic alignment, the
presence of linear polarization in the $J=0$ upper level to $J=1$
lower level transition at 1082.9 nm must be due to systematic errors. 
No atomic alignment is possible
with these quantum numbers, nor is it possible for levels involving
hyperfine structure of any $^3$He nuclei that might be present.  

In our figures, all data are taken from the dual-beam system, but we also
examined single-beam data.  
The signals in the two beams are very similar for all wavelengths 
outside of the helium lines, but that significant differences  are
present in the polarized helium profiles.  This is a clear sign of $I-(QUV)$ 
crosstalk in the helium lines.  The 
dual-beam corrections are clearly doing an excellent job 
at other wavelengths (for example, those in the \ion{Si}{1} line core).  
We surmise that the helium lines are evolving in intensity at least as
rapidly as the 1.2 s cadence of the modulation cycle, at the level of a few percent,
thereby producing spurious signals.   Faster modulation
seems appropriate for flare observations of the chromosphere. 
We defer further analysis of the flare kernel data for \ion{He}{1} to later 
work.

The noise levels of these Stokes $Q,U,V$ data are close to
$8\times10^{-4}I_C$.  The largest fringes remaining are in Stokes $V$ at the
level of $2\times10^{-3}I_C$.    The noise levels are more than adequate for
us to attempt inversions of the \ion{Si}{1} line data. 

\subsection{Quick look parameters}

We used the Stokes profiles to derive several simpler quantities.
From Stokes $I$ (intensity), we compute the $n=0,1,2$ moments
$M^{(n)}$ of the line profiles weighted by wavelength from line
center.  We define the continuum-subtracted line intensity as 
$$
I^\prime_x = I_C - I_x,
$$
where the Doppler shift $x$ is defined as  $c (
\lambda/\lambda_{LAB} -1)$ with $c$ in \velu{}. Then we define
$$
M^{(n)}= \int I^\prime_x x^n dx.
$$ The ``Doppler shift'' of the line is $v=M^{(1)}/M^{(0)}$ \velu, the
line width (not shown here) is $w= \sqrt{M^{(2)}/M^{(0)}}$ \velu.  We also computed
``quick-look'' quantities from the $IQUV$ Stokes parameters.  These include a LOS
``magnetogram'' which is simply the median of the ratio of Stokes $V$
to the first derivative of Stokes $I$ with respect to wavelength over wavelengths
of significant line absorption or emission.  The other magnetic parameter is
the field ``azimuth'' which is $\frac{1}{2} \arctan (U/Q)$ where again
median values of the ratio $U/Q$ are taken across both the \ion{Si}{1} line
and \ion{He}{1} multiplet.  The relationship of these quantities to physical
parameters in the Sun arises only when the polarization is dominated
by the Zeeman effect \citep[see,
  e.g.][]{Jefferies+Lites+Skumanich1989} and only when the
polarization $Q^2+U^2+V^2 \ll 1$, and when the Stokes profiles
all originate from the same physical volumes underlying a given pixel. 
Nevertheless, such quantities as
LOS magnetograms are very familiar to solar physicists which, with
care, illustrate properties of the solar magnetic field.

Figures~\pref{fig:evolone} and ~\pref{fig:evoltwo} show 
some of the quick look parameters and
continuum intensity from the three scans obtained before, during and after the impulsive phase.  
These we label phases ``bef'',``dur'' and ``aft''. 
The
white contours of Figure~\pref{fig:context} are from the Si core data
shown in the second row of Figure~\pref{fig:evolone}.  
Figure~\pref{fig:evoltwo} shows parameters related to the magnetic field. 
Salient features
of these plots include:

\begin{itemize}
\item{} The IR continuum shows only a weak brightening during scan ``dur''.
\item{} Both the photospheric\footnote{Other photospheric lines, not shown, 
also show emission cores in the FIRS flare footpoint spectra:  1081.83 nm (\ion{Fe}{1}), 1083.91 nm (\ion{Ca}{1}), 1084.40 nm (\ion{Si}{1}?).}  \ion{Si}{1}
 and chromospheric \ion{He}{1} lines show considerable brightening during scan ``dur'', in the line cores. 
\item{} The filament seen in the \ion{He}{1} line core images, lying roughly 
along the neutral line seen in ``bef'' scan magnetograms, disappears 
by scan ``aft''.  
\item{} The \ion{He}{1} absorbers in the filament are seen moving upwards by between 5 and 10 \velu{} in 
scan ``bef''.  During scan ``aft'', the filament is replaced by
 a diffuse region of \ion{He}{1} line absorption. 
\item{} Magnetograms show only subtle changes from scans ``bef'' to ``dur'' and ``aft''.
\item{} The photospheric magnetic azimuthal angles show systematic changes as the flare 
progresses. 
\item{} The \ion{He}{1} $U/Q$ signals during scan ``dur'' 
have coherent structure whose origin includes at least some 
$I\rightarrow (QU)$ crosstalk.  The spectral profiles show that 
they certainly cannot be interpreted as 
a traditional Zeeman-induced linear polarization.  
\end{itemize}

During scan ``dur'' (the impulsive phase), magnetic quick-look data for the \ion{He}{1}
line cannot be trusted to represent magnetic fields because of cross-talk. 
We will attempt to correct for this crosstalk in a future publication, since even
in the presence of atomic polarization, the 1083 nm multiplet still retains 
accurate information on field azimuthal angles, for field strengths $> 10$G (the strong field
limit of the Hanle effect). 
To quantify the magnetic field changes we
turn to inversions for the \ion{Si}{1} $IQUV$ data. 

First we will use properties of the \ion{Si}{1} line and continuum to
constrain the depth of penetration of flare energy that is sufficient
to change the temperature structure in the deep chromosphere and
photosphere.  At a first glance the continuum intensity appears to
change very little, if at all.  But both p modes and granulation
modify the continuum intensity at a level of a few percent at angular
resolutions similar to those of FIRS
\citep[e.g.][]{Sanchez-Cuberes+others2000}, making relatively small
changes difficult to see in slit rasters.  Careful examination of the
FIRS spectra obtained beginning at 17:46:16 UT, show a significant
brightening of $4\pm1$\% above levels in the neighboring spectra taken
13 seconds before and after, close to the flare kernel observed in
RHESSI and line data.  Evidence for this is shown in
Figure~\pref{fig:continuum}.  These data include variations in the
transparency and seeing conditions of the atmosphere and hence vary
significantly from row to row in the figure. As is obvious in the
figure, transparency variations were strongest in the first scan,
getting progressively weaker in the second and third scans.  However,
relative intensities {\em along} each row in each panel can be fairly
compared.  The uncertainty quoted above is a $1\sigma$
statistical variation of the detrended intensity measured along the
rows immediately adjacent to the row containing the flare. 

The coherent streak of brightness in
continuum data from 17:46:16 has all the
characteristics of a genuine brightness increase associated with a
white light flare.  This picture is supported by HMI continuum data
from the SDO spacecraft, which shows a $\approx$5\% increase in continuum
intensity during the flare.  Its appearance only in one FIRS scan indicates
a very rapid evolution, characteristic of an origin from dense
photospheric material which has a radiative relaxation time of 1-2
seconds \citep{Spiegel1957}.

\section{Analysis}
\label{sec:discussion}

The spatio-temporal behavior of the flare as obtained by FIRS is
summarized in Figures~\pref{fig:evolone} and ~\pref{fig:evoltwo}.
Remarkably, the slit happened to scan across the flare footpoint
ribbons at the flare peak, 17:46 UT (Table~1).  Also shown on the figure is the
core of the flare-related acoustic source \citep{Donea+others2014}.  
It is clear that FIRS managed to capture those locations in
solar-$y$ heliographic coordinate that correspond to the time and
place of the acoustic source.

\subsection{Helioseismic holography}

Seismic transients from solar flares can be detected by pre-processing
solar data and applying the analytical technique of helioseismic
holography to Doppler measurement of the active region hosting the
solar flare.  \citet{Donea+others2014} analyzed Doppler maps from the
Helioseismic and Magnetic Imager instrument (HMI;
\citealp{Schou+others2012}) on board the Solar Dynamics Observatory
satellite (SDO). HMI measures properties of photospheric dynamics and
magnetic fields every 45 seconds.  \citet{Donea+others2014} generated
Postel projection maps of the seismic emission of NOAA 12017.  We
refer discussion of the principles of seismic holography to Section 4
of \citet{Lindsey+Braun2000}, with application to flare observations
to \citet{Donea+others1999}.  Briefly, the seismic responses to the
flare perturbations are identified through an excess of the emission
power, $|H_{+}(r, t)|^{2}$.  Each pixel in an image of $|H_{+}(r,
t)|^{2}$ is a representation of the coherent acoustic power for waves
that have propagated downward from the focus, traveled thousands of
kilometers beneath the solar surface, and re-emerged into a pupil a
significant distance from the focus. With this technique
\citet{Donea+others2014} uncovered a weak but significant seismic
source at the footpoint shown in Figure~\pref{fig:context}.
Hard X-ray emission, magnetic transients and strong UV footpoint
emission were analyzed by \citet{Donea+others2014}, confirming that
the seismic source is indeed associated with the flare.  

\subsection{The depth of penetration of flare energy}

By comparing the brightness of models of continuum and \ion{Si}{1} 1082.7 nm
line to observations, we can in principle constrain the depth in the
atmosphere to which significant heating from above can penetrate.  Our
approach is simple.  We ask: what are the deepest and shallowest
layers in the atmosphere heated by the flare that are compatible with
the data?

To preface the model calculations below, we note that the flare
\ion{Si}{1} profile (Figure~\pref{fig:stokes}) resembles classical
\ion{Ca}{2} $H$ and $K$ self-reversed profiles
\citep{Linsky+Avrett1970}, but with far weaker line absorption wings.
The \ion{Ca}{2} line cores, much more opaque than the line of silicon,
form in the chromosphere with a source function dominated by
scattering.  The simple observation of a self-reversed profile of
\ion{Si}{1} implies a significant column mass, much higher than that
for the calcium line.  The breadth of a Doppler-broadened,
self-reversed line is larger than an optically thin line formed under
the same conditions by the factor $\approx \sqrt{\ln \tau_0}$ where
$\tau_0$ is the line center optical depth.  For $\tau_0=100$ this
factor is over 2.1.  The self-reversal is also very narrow (FWHM $\approx
0.015$ nm, see Figure~\pref{fig:stokes}), indicating turbulent speeds
of FWHM/1.66 $\approx2.5$ \velu{} where the
line core forms.  A profile averaged along the region with
obvious \ion{Si}{1} emission in Figure~\pref{fig:stokes} is shown in
Figure~\pref{fig:nlte}.  The averaging washes out the self-reversal in
the latter plot.

We model the \ion{Si}{1} line and the neighboring IR
continuum both during and outside of the impulsive phase.
We performed nLTE radiative transfer calculations, in several 1D
models of the solar atmosphere, following the tradition of
\citet[][henceforth VAL81]{Vernazza+Avrett+Loeser1973,
  Vernazza+Avrett+Loeser1976,Vernazza+Avrett+Loeser1981}.  We solved
nLTE statistical equilibrium equations for atoms of H, C, Si and Fe using the program RH
\citep{Uitenbroek2000}.  These atoms were chosen because UV radiation 
controlling the \ion{Si}{1} spectral line at 1082.7 nm is dependent on the nLTE 
solutions of these abundant elements.
We considered using 
one of several flare models \citep[e.g.][]{Machado+Emslie+Avrett1989}.
However, these models were constructed to try to identify the origin of white light 
emission in flares.   Our goal is different, to try to see if modeling can provide 
a depth of penetration of flare energy into the photosphere.  
Therefore we adopted a different, more straightforward strategy.
We started with the model ``C'' of VAL81 and
explored the effects of introducing
temperatures plateaus of the form
$$
T_e = T_0 - T'_1 \log m,   \ \ \ m_2 > m, 
$$ 
where $m$ is the column mass of the atmosphere, $T_0,T'_1$ are
non-negative constants, and $m_2$ is a column mass above which 
temperatures are changed.  We have three free parameters, 
and so our results will not be unique.  But 
such plateaus, with small gradients $T'_1$, 
have justification at least during some phases of flaring plasmas 
seen in radiation hydrodynamic calculations \cite[see the 50s panel of Figure 3
  of][for example]{Allred+others2005}.  The main sensitivity of the emerging spectra
is to the two parameters $a$ and $m_2$.   Given an estimate of $m_2$ the 
height of the energy penetration follows from the $m(z)$ relationship for the model. 

We made calculations in two limits:
in the calculations shown in the Figures below we allowed the
atmosphere to relax to a state of hydrostatic equilibrium; in the
other limit we merely solved the statistical equilibrium equations with no
such adjustment.  The sound crossing time of the photosphere is on the
order of a few scale heights divided by 7 \velu{}, a minute or so,
comparable to the duration of the flare impulsive phase.  These limits
probably span the behavior of intensities from an evolving atmosphere.
The differences between the calculations are small in photospheric
layers but are significant for regions and spectra formed above 600
km above the photosphere.  Such differences do not affect our
conclusions which depend only on the photospheric \ion{Si}{1} line.

These calculations are not state-of-the art in terms of dynamics, our
focus is instead on a careful treatment of the formation of the \ion{Si}{1}
1082.7 nm line and of the continua formed between 125 and 180 nm for
later comparisons with SDO/AIA data.  We therefore took care to use
modern and complete atomic data for the Si and Fe neutrals.  We used
atomic energy levels and transition probabilities from NIST up to and
including the $4p$ levels in Si, and we used photoionization cross
sections from the OPACITY project \citep{Seaton1987}, treated as
outlined in \cite{Judge2007a}. Collisions with electrons were treated
using the impact approximation for permitted transitions
\citep{Seaton1962b}, Seaton's semi empirical formula for direct
ionization \citep{Allen1973}, and a collision strength of 0.1 for
forbidden transitions.

Figure~\pref{fig:nlte} shows, in the right panels, computed and
observed profiles of \ion{Si}{1} 1082.7 nm, with all intensities normalized to
quiet Sun values.  These calculations are representative of two
limits of the value of $m_2$ --  and hence height of penetration -- used in the 
models.  

The first class (upper panel) allows penetration 
of energy and enhanced temperatures down to photospheric layers - we 
allowed temperatures to rise down to 0 km height by adding various 
plateaus at such depths.  
Remarkably, the model shown produces an acceptable match to the observed profiles and
continuum (the \ion{He}{1} line is not modeled here).  Exploring
different temperature plateaus we 
determined that a reasonable agreement with the line and continuum 
observations 
requires the flare energy to penetrate and heat down to a height of 
$\gta 100 \pm 100 $ km above the photosphere.  The ``error bar'' comes from
the need to produce the $4\%$ enhancement in continuum emission ($< 200$ km) with temperatures
that can match the \ion{Si}{1} profile, both features spanning the region between 0 and 700 km.

The second limiting case is one where flare energy penetrates only to the 
mid-upper chromosphere.  
Downward propagating radiation enhances the cores of lines, a typical
calculation is shown in the lower panels.  The line width is very narrow
even though we adopted non-thermal speeds (microturbulence) of up to 
8 km/s in the middle chromosphere (close to the sound speed). The continuum, 
formed predominantly in the photosphere with a tiny contribution from optically thin
emission in the plateau, is close to the pre-flare level.  
The computed continuum 
includes thermal photospheric emission as well as hydrogen
recombination from the plateaus. These two contributions have been 
discussed by \citet{Machado+Emslie+Avrett1989, Kerr+Fletcher2014}, among others.  
The contribution from the latter is small in our calculations, the Balmer continuum
originating from an optically thick layer near 350 km and the
longer-wavelength ($> 364$ nm) H$^-$ and Paschen continua near 0 km.

The core of the \ion{Si}{1} line during the flare is broad compared
with a thermal width near 1.8 \velu{} (Figure~\pref{fig:nlte}),
and like the well-studied \ion{Ca}{2} $H$ and $K$ lines the origin of
this width appears most naturally explained through scattering (see above).  Some
decades ago there was a discussion of the Wilson-Bappu effect, an
empirical relationship between the width of the core of the
\ion{Ca}{2} lines and stellar luminosity, in favor of line formation
in terms of scattering \citep{Ayres1979} and not optically thin micro-
or macro-turbulence \citep{Fosbury1973}.  The presence of the narrow
self-reversed core seems irrefutable evidence for the presence of
scattering and argues strongly for a deep formation of the core. Only
calculations of penetration of flare energy to the photosphere produce
lines broadened by scattering and self-reversals, the latter happen to
be weak in the case shown in Figure~\pref{fig:nlte}, but not in
obvious disagreement with the observed profile.

We stress that the detailed structure of our calculations is not
unique and should only be viewed as an attempt to find the depth of
penetration of significant heating during the impulsive phase of the
flare.  Overall, our comparisons with observations of the \ion{Si}{1} 1082.7
nm line, and taking into consideration the difficulties of tying down
the continuum intensity during the flare, we conclude that {\em
  heating sufficient to change detectably the photospheric temperature
  occurs at least to about 100 $\pm 100$ km above the visible
  photosphere}.  Based on 
an exploration of values of $T_0,m_2$ in our model, 
we believe that this aspect of our calculations is robust.

\subsection{Inversions}

We used the code MELANIE \citep{Socas-Navarro2003} to invert the \ion{Si}{1}
Stokes $IQUV$ profiles to derive the vector magnetic field in the
photosphere. This was done only for scans obtained before and after the impulsive phase.
Codes exist for inversion of the \ion{He}{1} multiplet
\citep[e.g.][]{Lopez-Ariste+Casini2002,Lagg+others2004,Asensio+Trujillo+Landi2008}, but we
have not attempted such inversions yet 
because we must deal with significant crosstalk 
in the \ion{He}{1} $QU$ profiles during the flare, and because 
outside of the flare these profiles are mostly of low signal-to-noise ratio.

The observed \ion{Si}{1} line --
$3p4s~^3\!P^o_2 - 3p4p~^3\!P^e_2$ (lower and upper levels respectively) -- 
forms between $\approx$ 100 km (wings) and 600 km (core) above the 
photosphere in our 1D models.  
MELANIE solves for a solution to the Milne-Eddington
equations (source function linear with optical depth) for lines with
Zeeman-induced polarization, minimizing differences between observed
and computed profiles. The solution includes the vector
magnetic field (with its 180$^\circ$ ambiguity), 
opacity, Doppler width and shift, damping parameter, non-magnetic
filling factor.  The Milne-Eddington approximation is a simplification
that surely is invalid during the flare itself.  But before and after the flare 
its use appears reasonable, outside of bright flare ribbons and below
say 600 km in the atmosphere.  Our conclusions will be based only
on the non-flaring atmosphere. 

We inverted all five scans.  We set the statistical uncertainty of
each data point to $10^{-3}I_C$ to evaluate values of $\chi^2$,
estimated using the measured fluctuations in $QUV$ at typical
continuum wavelengths.  For comparison, some of the best vector
polarimetric data, the ``deep mode high S/N'' observations from the SP
instrument on the Hinode spacecraft have rms noise of $3\times10^{-4}I_C$ in
the 630 nm region, for integrations of 67 s \citep{Lites+others2008}.
Outside of the flare scan, the distribution of $\chi^2$ peaks near 40,
showing that systematic errors are large and/or the model
parameterization is poor.  Given the residual fringing and other
artifacts evident in the data, this does not by necessity imply that
the model is poor.  The reproducibility of the inversions was tested
by initializing the same dataset with two different random initial
guesses.  
The resulting rms variations in the magnetic field strength
$B$ are 140 G, inclination 18$^\circ$, azimuth $41^\circ$, and the LOS
$B$ 30 G.

Figures~\pref{fig:invone} and \pref{fig:invtwo} show results of
inversions of the scans obtained before and after the flare, begun at
16:29:26, 16:55:58, and 18:01:55 UT.  No attempt at a resolution of
the 180$^\circ$ ambiguity in the field azimuth has been made, and
the angles are defined relative to the local vertical\footnote{The vertical
direction of center of the region is rotated 39.6$^\circ$ E-W and 15.8$^\circ$ S-N
relative to the LOS.} (inclination) and in the plane of sky 
(azimuth, zero and $180^\circ$ being along the E-W direction). 
Circles show the location of the center of the acoustic source.
Figure~\pref{fig:invone} shows measured changes in magnetic parameters
in the two scans obtained before the flare.  There are detectable
differences across the bulk of the field of view in all magnetic
parameters.  Focusing on data in the circled region of the acoustic
source, we see a significant increase in the field strength in this
region, accompanied by becoming more inclined to the vertical
direction (data shown in the first two rows of the Figure).  
Note that the circled region is some 5$\arcsec$ from the polarity inversion line.
The maps of $B$ suggest that a channel of weak field is moved to the west 
by an arcsecond.  
Initially
the field is inclined at some 130$^\circ$ to the vertical. But by 17:05 UT two bands 
of field connected in a ``Y-shape'' on its side in the image appear more inclined to 
the vertical. 
The field azimuth in the ``Y'' shape departs significantly from initially E-W to
more N-S.  
The LOS field within at the circle's center shows an increase that results from 
increases in $B$ despite the decrease in inclination.  It is unclear from our data 
if these changing fields arise from motions of field vertically 
(flux emergence) or horizontally (flows).   There is little 
evidence for vertical motions from the LOS velocity measurements shown in
Figure~\pref{fig:evolone}, but the inversion data (not shown) reveal a 
very small (-0.3 \velu{}) blue-shift pattern in the \ion{Si}{1} data in the 10:55:58 UT 
scan that might conceivably be associated with the upper part only 
of the ``Y'' pattern seen in the magnetic data.   

The scans upon which the inversions are based are 26 minutes apart.
The above changes are unremarkable when compared to the larger field
of view, except that they are within the circle encompassing the
acoustic source and they are significant in all magnetic parameters.

Figure~\pref{fig:invtwo} shows measured changes before and after the flare itself, 
scans begun 66 minutes apart.  The difference panels show again an increase 
of $B$ and azimuth, and a weak reduction of inclination, in a band in the E-W direction
cutting through the circled region, flanked by regions of increased 
inclination just to the S and N.  This sheared region (differentially 
changing field inclinations with time) 
appears aligned with the bright 
footpoint emission seen in the core intensities 
of the \ion{Si}{1} and \ion{He}{1} lines.  The data are noisy, however.

Thus, our analysis hints that magnetic fields 
associated with the particular acoustic source evolve to become more sheared 
(i.e. inclination angles diverging in time), 
stronger (perhaps due to flux emergence) and rotated relative to the EW direction,
during the flare.  These results appear to correspond to a mixture of earlier results.
\citet{Wang+Deng+Liu2012} found penumbral fields which became more vertical 
after flaring. 
In contrast, for some flares 
\citet{Martinez+others2008,Wang+others2012} reported field lines
highly inclined to the vertical after a flare-associated 
seismic transient.  We note that the seismic source we have analyzed 
is unusual. It is found near a magnetic pore, emerging from a magnetically quieter 
area somewhere between the main two
sunspots of the AR12017 \citep{Donea+others2014}, instead of in a penumbra.

Lastly, if flux emergence were responsible for 
these measured changes in magnetic field, 
in 1 hour the plasma and magnetic field moving vertically through the 
compressible sub-photosphere with a surface velocity 
$\lta 0.3$ \velu{} could have emerged from depths
no deeper than $\approx 200$ km. If advected by granules with 1 \velu{}
speeds, the flux could have emerged from no deeper than 600km. It is interesting to
consider how
such changes to the immediate subsurface structure might or might not affect the generation
of sunquakes.

\subsection{The mode of transfer of  flare energy down through the atmosphere}

Armed with a unique dataset, we have studied the depth of penetration
of flare energy down into the solar photosphere.  We have shown that
the detected changes in {\em thermal} structure in the atmosphere
reach the photospheric level, but barely.  Here we examine possible
modes by which the energy might be transported through the photosphere
into the deeper solar layers, thereby exciting the sunquake. 

The power in the main kernel of the acoustic source 
measured using seismology from HMI is \citep{Donea+others2014}: 
$$
P = 1.3 \pm 0.05 \times 10^{26} {\rm  erg~s^{-1}}. 
$$ 
This power is distributed over an area including the main kernel
centered at $(X,Y)=(518,264)$ (see Figure~\pref{fig:context}), the
source just to the SW requiring an additional $1.0\times10^{26}$
erg~s$^{-1}$.  The main source's spatial distribution is nearly
bi-Gaussian with a geometric mean full width at half-maximum (FWHM) of
$w=4.2$ HMI pixels, $w\equiv1.5\times 10^8$ cm.  The peak of the power per
unit area is $ F = P/ ( A=\pi a^2) $ with $a=w / 2\sqrt{\ln 2}$, or
$$
F=5\times10^9 {\rm  erg~cm^{-2}~s^{-1}}. 
$$
This should be regarded as a lower limit since both the holographic technique
and HMI have non-negligible angular resolutions.  The area $A=
2.6\times10^{16}$ cm$^2$ is strictly an upper limit for the same reasons. 

Let us consider first ``non-magnetic mechanisms'' by which energy is transported 
to the acoustic source.  In this picture the
changing magnetic field generates thermal perturbations 
indirectly via the end product of large coronal magnetic restructuring
(conduction, particles, local downward radiative heating),
channeling some flare energy
into the photosphere.   We can estimate energy fluxes into the acoustic 
source region that
are compatible with our observations in several ways.  First, we note that the 
excess thermal energy radiated from the photosphere during the few minutes 
of the rise phase is roughly 4-5\% (i.e. the measured continuum enhancement) 
of the unperturbed solar radiative flux density $F_\odot = 6.33\times10^{10}$ \flxu: 
$$
P_{RAD} \approx  0.04 F_\odot A \lta 7\times10^{25} {\rm \  erg~s^{-1}}.
$$
The radiative cooling time of photospheric plasma is 1-2 s
\citep{Spiegel1957}. Curiously then, although $P_{RAD} \sim P$, this excess thermal 
energy is
simply radiated into space on such timescales, and is unavailable to contribute to $P$.  
We can look at the {\em enthalpy} flux
$F_{enth}$ associated with bulk flows into the photosphere, for this we need
a measurement of plasma motions and we turn to the \ion{Si}{1} line core emission
which forms between 200 and 500 km in our models.   We use pressures 
$p=2\times10^4$ dyne~cm$^{-2}$ 
and densities $\rho=4\times10^{-8}$ g~cm$^{-2}$. 
corresponding to 300 km height.  These are conservatively high values for 
average thermal properties of the plasma where this line is formed, favoring 
higher estimates of energy transport. 

A careful comparison of the flare emission core and the pre-flare absorption 
profile of the \ion{Si}{1} line 
reveals an upper limit to differential flows of roughly 0.5 wavelength 
pixels, 0.5 \velu{}.   This is equivalent to 2.5$\sigma$ 
where $\sigma$ is the sensitivity of 
the Doppler shifts from our FIRS spectra. 
A Doppler photospheric signature of the flare is
present in HMI data at the location of the seismic source with 
a shift equivalent to $u \approx 0.3 - 0.5$ \velu{} \citep{Donea+others2014}.
However, such filtergram data, scanning wavelengths in time, cannot be 
trusted during flaring and so we adopt the upper limit above.  
We then find an upper limit to the 
enthalpy energy flux of
$$
F_{enth} \lta   \,\frac{5}{2}p \, u\, A \approx 6\times 10^{25} {\rm  erg~s^{-1}}.
$$ 
The close agreement of
the upper limit for $F_{enth}$ with $F_{rad}$ means that the excess
energy radiated by the photosphere during the flare can be supplied by
a bulk flow of energy associated with a subsonic downflow of
0.5 \velu{} induced (somehow) by the flare.  The power in the acoustic
pulse is a factor of at least 2 larger than our optimistic estimate of 
$F_{enth}$.


If however the pressure pulse involves high frequency phenomena ($\nu
> c_S/H \approx 60$ mHz, where $H$ is the pressure scale height and
$c_S$ the sound speed), the pulse would be invisible to observation
except as a broadening of spectral lines to at most the sound
speed (for linear waves), the lines being formed over a length
$\approx H$ in a stratified atmosphere.  The WKB expression for the
energy flux density (propagating both upwards and downwards) at the sound
speed is
$$
F_{wave} = \rho c_S \langle \xi^2 \rangle  
\ \ \ {\rm  erg~cm^{-2}s^{-1}},
$$
where $\xi$ is the velocity amplitude of the wave.  We can set limits on 
$\xi$ through the measured line broadening and line profiles during the flare
itself.  Before and after the flare, the inversions yield 
$\xi \lta 2.8$ \velu{}.  During the flare 
the measured line wings are similar in shape to the pre- and post- flare profiles. 
The 
emission core of the profile has a FWHM of 
0.05 nm (Figures~\pref{fig:stokes} and \pref{fig:nlte}). 
Treated as optically thin emission, this FWHM is 
equivalent to an e-folding Doppler broadening speed of 
$8$ \velu{}.
In the presence of scattering this is a strict upper limit. 
To estimate the energy flux available in such modes
we again use $\rho = 4\times 10^{-8}$ g~cm$^{-2}$, 
and the upper limit of 8 \velu{}.  Assuming that only half of the 
waves are emitted downwards, we find
$$
F_{wave} A \lta 2.5\times10^{26}\ \ {\rm  erg~s^{-1}}.  
$$
But this, we believe, is a gross over-estimate.  
Firstly, 
the scattering leads to 
emission profiles a factor of 
$\approx \sqrt{\ln \tau_0}$ broader than mere Doppler broadening
where $\tau_0$ is the line center optical depth.   We are able to reproduce the 
core \ion{Si}{1} emission using microturbulent speeds of 1-2 \velu{}
and the full Voigt profile, reducing the above 
estimate $F_{wave}$ by a factor of at least 16!
Secondly, the line profile 
shows, within a broad 
emission core, a very narrow self-reversal during the flare 
(lower left panel of Figure~\pref{fig:stokes}), indicating both 
scattering-induced line broadened profiles and 
values of $\xi$ in the line core 
far smaller than those adopted above.
Lastly,  any high 
frequency waves with frequencies of a few Hz or less 
are rapidly damped in the photosphere by the continuum radiative exchange processes 
first modeled by \citet{Spiegel1957}.
The thermal perturbations associated with 
high frequency wave energy rapidly radiate this energy 
from the photosphere on a timescale of 1-2 s.  
Only waves with frequencies in excess of several Hz could propagate 
down into the interior unmodified by radiation damping. All things considered, 
it seems unlikely that the power of the sunquake can be provided by 
such high frequency waves.

{\em We conclude that 
non-magnetic modes of energy transport into the interior are very unlikely to
be sound waves}.  More likely is a coherent downward-moving plug of plasma carrying enthalpy of
almost the right magnitude, but we have above already set an upper limit to this process 
that is optimistically a factor of two smaller than needed.

Consider in turn the energetics of the Lorentz force picture.  The
field strength from inversions from the \ion{Si}{1} line, before and
after the flare, is of order 800 G from which the magnetic energy
density is $B^2/8\pi = 2.5\times10^4$ erg~cm$^{-3}$, about 1/3 of the
photospheric thermal energy density, $\frac{3}{2} p_{ph}$ (the latter
is a lower limit since we neglect latent heat of ionization).  The
Alfv\'en speed $c_A$ for a photospheric density of $2.6\times10^{-7}$
g~cm$^{-3}$ is 4.3 \velu{}, so the local magnetic energy flux is at
most
$$
F_M A \lta c_A \frac{B^2}{8\pi} A = 2.4\times10^{26} \ \ {\rm erg~s^{-1}},
$$ 
scaling as $B^3/\sqrt{\rho}$.  This estimate is entirely a local
one, it does not take into account the connections of the magnetic
field throughout the entire flux system and the fact that momentum and
energy is readily imparted to localities from a far larger reservoir
of magnetic energy encompassing the entire volume of the active region.  Thus there is
naturally sufficient energy in the magnetic field of the entire active region 
to account for the
acoustic source.  But it should be remembered also that 
only a fraction of the 
total magnetic energy is available as free energy, 
only a fraction will penetrate into the interior, and 
measured changes in magnetic fields before and after flares 
are small relative to the ambient field. Estimating the free energy
change over the entire active region is not a simple task, fraught with 
problems  \citep{Derosa+others2009} and so is not attempted here.

Instead let us assume that a magnetic force is responsible for the impulse
into and below the photosphere at the acoustic source.  The same
argument for the enthalpy flux applies here no matter if a Lorentz-forced flow
is responsible, and the same inadequate energy flux
results from the stringent limit on flows found from the spectra of
the \ion{Si}{1} line.  The reason is simple: magnetic fields are 
frozen to plasma under conditions of high magnetic Reynolds number, 
motions of plasma {\em must} accompany any forcing no matter its origin. 
The collision time for a proton to exchange 
its momentum with a hydrogen atom is of order 
$4\times10^{9} / \sqrt{T} n_H$ sec. With $T=6000$ K, $n_H \approx 10^{17}$ cm$^{-3}$, this
time is far smaller than any dynamical time scale
\citep{Gilbert+Hansteen+Holzer2002}. The partially ionized plasma behaves
dynamically as a fully ionized plasma with the same charged particles 
but with the additional neutral mass.  
Again therefore we must appeal to unresolved motions, i.e. waves.  But
the fastest magnetosonic mode is the field-aligned modified sound wave
when the plasma $\beta > 1$, so that the same conclusions are drawn
as for the non-magnetic case.
{\em The magnetic forces are either (a) incompatible with the
  observations of line profiles revealing down-flowing material with
  insufficient energy flux to account for the acoustic source, or (b)
  are in the form of modified high frequency sound waves with the same
  problems that the sound waves have.}

\section{Discussion and Conclusions}

We have reported some unusual spectral and polarimetric profiles of lines
of \ion{Si}{1} and \ion{He}{1} 
obtained at a flare footpoint at infrared wavelengths.  Several
photospheric lines revealed line core emission in excess of
non-flaring conditions, and the helium 1083 nm multiplet almost
doubled in brightness.  Using the enhanced brightness of the
\ion{Si}{1} 1082.7 nm line and the neighboring continuum, we have
demonstrated via 1D radiative transfer models that flare-related
heating can be detected all the way to the photosphere, to 100 $\pm
100$ km.  This is deeper by several scale heights than the expected
depth of penetration of hard X ray emission.  Our models merely show
the depth to which heating is observed to occur through the flare
mechanism(s), they shed no light on the nature of this transport from
the corona into the Sun's deeper atmosphere.  It could be a direct
mechanical effect or a two-stage mechanism of mechanical transport
followed by radiative back warming \citep{Machado+Emslie+Avrett1989}.
Using dynamical models in a stratified atmosphere,
\citet{Allred+others2005} found the flare energy penetrated only to
the mid chromosphere (800 km).  \citet{Martinez+others2012} found
heights of 305 $\pm$170 km and $195 \pm 70$ km, respectively, for the
centroids of the hard X ray and white light footpoint sources of a
flare observed stereoscopically.

How robust is our new result?  We can produce significant emission in the
core of the \ion{Si}{1} line using both deep and shallow penetration models (enhanced
temperatures only above 500 km). But the shallow models can be eliminated:
the continuum intensities are unchanged from pre-flare conditions; the
\ion{Si}{1} line cores are far too narrow; also, only models with energy
penetrating down into the photosphere have sufficient opacity to
produce the opacity broadening and subtle narrow 
self-reversal observed in the very center of the
line.  The combination of broad, self reversed line emission and a
brighter continuum appears to be a clear signature of flare heating down
to $100\pm100$ km.  This picture also eliminates the possibility that
hydrogen recombination radiation contributes to the visible and infrared 
continuum emission
for this flare \citep{Kerr+Fletcher2014}.
One should expect that our conclusions  depend strongly on the adopted
model of the photosphere/chromosphere: surely 1D models are inadequate 
for studies of the Sun's atmosphere in general? It is sometimes forgotten
that the photosphere/low chromosphere are characterized 
by subsonic motion, and that therefore the atmosphere is strongly
stratified\footnote{Very dynamic phenomena seen at UV wavelengths, for example with
the HRTS or IRIS instruments, are formed 
mostly in less dense structures above the 
stratified layer that is optically thick to most UV radiation.}.  
It is this essential property of these plasmas, 
in which the \ion{Si}{1} diagnostic line we have used is formed, 
that is most critical for determining the depth of penetration
of flare energy.   There appears to be no credible way to reconcile the 
salient line profile and continuum observations with heating 
occurring purely above 200 km.   Whatever model we would choose to
use, we would still require continuum and line 
formation  within the photosphere, not the chromosphere. 

We found significant pre- and post-flare changes in the magnetic
conditions within and outside of the acoustic source region.  Within
the source region, these include an increase in magnetic field
strength, a rotation of the magnetic azimuth and a reduction of
inclination with respect to the solar vertical.   The further interpretation
of these changes, part of the evolution of the entire active region,
is beyond the scope of this paper. 

We detected no signature of downward energy transport capable of
carrying $2.3\times10^{26}$ erg~s$^{-1}$ 
needed to account for the
acoustic source, that is compatible with our ground-based
observations.  Curiously, both line and continuum emission is present
along an extended ribbon, but the acoustic emission is confined to a
smaller region only associated with the brightest parts of the ribbon.
At some future time, the flux of energy in {\em something} propagating
downwards through the Sun's atmosphere must be detected.  We have
shown using current spectroscopic capabilities that macroscopic
motions driven magnetically or otherwise carry insufficient energy by
at least an order of magnitude, and that high frequency acoustic and
magnetic waves are also are very likely to fall short.
Spectropolarimetric analysis reveals magnetic fields that are too
small at the acoustic source for magnetic wave modes to carry the
energy ($c_A \lta c_S$), and if a Lorentz force is responsible for the
impulse, we must at some point also observe either a Doppler shifted
downflow of plasma or lines broadened enough for the flux $\rho c_A
\xi^2$ to carry sufficient energy.  If we have captured the essential
dynamics of this particular flare in our observations, it seems that
we have ruled out two important classes of models for the forcing of
acoustic sources under the photosphere.  Anything that relies on
transport of energy (radiation, conduction) more than momentum seems
to be
incompatible with our data.

Our work therefore raises a problem, namely that the energy is
transported downwards in a fashion that is somehow invisible to our
observations.  Yet the latter include the infrared continuum and lines
of \ion{Si}{1} and \ion{He}{1} and they span the entire photosphere
and chromosphere of the Sun.  Is it possible that we have missed the
region of ``action''?  We do not believe so since our slit passed
right across the bright flare kernel accompanying the acoustic source
at the right time.  Perhaps the energy is transmitted in intense but
very short ($< 1$ s) bursts which, when integrated over the 13 s dwell
time at each slit position, might contribute only to very broad
spectral lines that are washed out spectrally and/or temporally in our data.
Such variations might be absent in the continuum and line emission
that we can measure.  Or perhaps the energy is transmitted to the source
over a larger area of the visible surface, and focused somehow into
the source itself.  Our estimate of $A\lta 2.6\times10^{16}$ cm$^2$
is our best estimate based upon the acoustic wave analysis of
\citet{Donea+others2014}.  Other alternatives might appeal to high
energy particles accelerated in the corona.
The power delivered by electrons of energy sufficient to reach
altitudes of 100~km above the photosphere can be estimated using the
RHESSI quick-look spectral fits to obtain photon spectral index and
intensity, which is related straightforwardly to electron power using
the collisional thick target approximation \citep[see
  e.g.][]{Fletcher+others2007} and the approximation that to traverse
a column of $N \rm{cm}^{-2}$ requires an electron of energy $E^2 =
N/(10^{17}\mu_o)$, where $\mu_o$ is the pitch-angle cosine of the
electron (set equal to 1 here) and $E$ is measured in keV.  In the
VAL-C model, 100~km above the photosphere corresponds to $N =
1.3\times 10^{24} \rm{cm}^{-2}$ (assuming a pure hydrogen target)
meaning that electrons of 3.6~MeV are required. The HXR spectrum is
hardest and most intense at 17:46:00-17:47:00, during which time the
photon spectral index at high energies is $\gamma = 3.23$, and the
intensity at 50 keV is approximately 1 photon $\rm{s^{-1} cm^{-2}
  keV^{-1}}$. Using these data in equations (1) and (2) from
\cite{Fletcher+others2007} gives a power in electrons above 3.6~keV of
$4.4\times 10^{24}{\rm erg~s^{-1}}$.  Thus the electrons appear to fall 
short by a factor of 30 from directly depositing the needed energy to the 
100 km depth.  We can of course speculate, along with many previous 
workers, that 
high energy protons might provide the needed energy flux deep in the 
solar atmosphere.  
Protons require energies $\sqrt{m_p/m_e}$ higher, i.e. in excess of 160 MeV. 
Such protons are sometimes seen in interplanetary space, associated with 
flares.  But to remain invisible to
our observations, such beams must penetrate and dump their energy and momentum {\em
below} the photosphere, bumping up the energy requirements through the column mass
by a factor of three or more.  
We see no evidence for an
up-welling of energy or plasma in response to such energy deposition below the
surface in our two scans obtained after the flare itself.

Our spectra serve to remind us of  difficulties in measuring magnetic and velocity 
field changes during flares.  The \ion{Si}{1} line profiles bear no resemblance to
what is usually assumed to ``invert'' photospheric lines, and 
the usual magnetograms and velocities 
render spurious results during a white-light flare. Such photospheric changes will 
similarly alter the formation of the \ion{Ni}{1} 676.8 nm or \ion{Fe}{1} 617.3 nm
and other lines routinely used on spacecraft.

In future, several promising lines of research should be
pursued for this uniquely well\--observed flare. We will 
analyze further the FIRS and IBIS data, exploring the
magnetic field changes across the atmosphere using lines formed at a
variety of depths from photosphere to chromosphere.  They have many
advantages over space-based data UV which tend to saturate during
flares, and over space based magnetic data owing to the limited modes
of operation and other characteristics of polarimeters in space.
Following the seminal calculations by \citet{Machado+Emslie+Avrett1989}, 
radiation hydrodynamic work along the lines of
\cite{Abbett+Hawley1999,Allred+others2005} is worth revisiting to help
further determine the nature of the heating mechanism above and beyond
the work presented here.

\acknowledgments We thank the observers at the DST for their help.  Sarah Jaeggli and 
Tom Schad provided software and advice for reducing the FIRS data used here. 
PGJ acknowledges discussions with Roberto Casini, Bruce Lites and Alfred de Wijn.
Part of this work was carried out under NASA grant NNX13AI63G.  The referee provided 
helpful comments on the manuscript.

\tabone

\figcontext

\figstokes

\figevolone

\figevoltwo

\figcont

\fignlte




\figinvone
\figinvtwo

\end{document}